\documentclass[letterpaper, 10 pt, conference]{ieeeconf}  % Comment this line out if you need a4paper
\usepackage{graphicx}
\usepackage{graphics} % for pdf, bitmapped graphics files
\usepackage{epsfig} % for postscript graphics files
\usepackage{mathptmx} % assumes new font selection scheme installed
\usepackage{times} % assumes new font selection scheme installed
\usepackage{amsmath} % assumes amsmath package installed
\usepackage{amssymb}  % assumes amsmath package installed
\usepackage{hyperref}
\usepackage{multirow}
\usepackage[table]{xcolor}% http://ctan.org/pkg/xcolor
\usepackage{cite}
\usepackage{longtable}
\usepackage{booktabs}
\usepackage{wrapfig,lipsum}
\usepackage{xtab}
\usepackage{caption}
\IEEEoverridecommandlockouts                              % This command is only needed if 
\title{\LARGE \bf React to This (RTT): A Nonverbal Turing Test for Embodied AI}

\author{Chuxuan Zhang$^{*}$,  Yasaman Etesam$^{*}$ and Angelica Lim% 
\thanks{$^{*}$Equal contribution. Authors are with the School of Computing Science, Simon Fraser University, 8888 University Dr., Burnaby, Canada
        {\tt\small \{chuxuan\_zhang,yetesam,angelica\}@sfu.ca}}%
}

\begin{document}

\maketitle
\thispagestyle{empty}
\pagestyle{empty}

% \begin{figure*}[ht]
%     \centering
%     \includegraphics[width = 0.9\textwidth]{Figures/characters.png}
%     \label{fig:enter-label}
%     \caption{Participants were asked to test six interactive virtual characters physically, emotionally and socially}
% \end{figure*}
% How do people use their faces and bodies to test the interactive abilities of a robot? Making lively, believable agents is often seen as a goal for robots and virtual agents but believability can easily break down. In this Wizard-of-Oz (WoZ) study, we observed 1169 nonverbal interactions between 20 participants and 6 types of agents. We collected the nonverbal behaviors participants used to challenge the characters physically, emotionally, and socially. The participants interacted freely with humanoid and non-humanoid forms: a robot, a human, a penguin, a pufferfish, a banana, and a toilet. We present a human behavior codebook of 188 unique nonverbal behaviors used by humans to test the virtual characters. The insights and design strategies drawn from video observations aim to help build more 
We propose an approach to test embodied AI agents for interaction awareness and believability, particularly in scenarios where humans push them to their limits. Turing introduced the Imitation Game as a way to explore the question: ``Can machines think?'' The Total Turing Test later expanded this concept beyond purely verbal communication, incorporating perceptual and physical interaction. Building on this, we propose a new guiding question: ``Can machines react?'' and introduce the React to This (RTT) test for nonverbal behaviors, presenting results from an initial experiment. 

% to evaluate a machine's capacity to respond meaningfully and believably in dynamic, real-world interactions.
% believability of interactive virtual characters 
% but breaks down in open environments
% as humans test the interaction capabilities of the agents
% what should characters react to

\section{Introduction}
% In 1950, Turing~\cite{turing2009computing} introduce the ``imitation game'' to answer the question ``can machines think?'' After that many tried to pass the test~\cite{shieber1994lessons}. One of the earlies works that showed how superficial language mimicry can fool people briefly was
% \textit{ELIZA}~\cite{weizenbaum1966eliza} developed in 1965. ELIZA respond to the input rougly as would a certain tain psychotherapists (Rogerians) do. However, for later competetion expectations raised by having the judges interacting with ELIZA beforehand. \textit{``For example, one were to tell a psychiatrist ``I went for a long boat ride.'' and he responded ``tell me about boats'' , one would not assume that he know nothing about the boats}.
In 1950, Turing~\cite{turing2009computing} proposed the ``imitation game'' as a way to address the question: ``Can machines think?'' Since then, numerous attempts have been made to pass this test~\cite{shieber1994lessons}. One of the earliest systems to highlight how surface-level language mimicry could  deceive users was ELIZA~\cite{weizenbaum1966eliza}, developed in 1965.
ELIZA simulated the conversational style of a Rogerian psychotherapist by reflecting users’ inputs in a seemingly meaningful way. 
% As Weizenbaum pointed out, ``If one were to tell a psychiatrist \textit{I went for a long boat ride}, and he replied \textit{Tell me about boats}..." one would accept this as normal~\cite{weizenbaum1966eliza}. 
However, expectations around such systems evolved. 
In 2014, a chatbot named Eugene Goostman, which portrayed a 13-year-old Ukrainian boy, was claimed to have passed the Turing Test during an event hosted by the University of Reading\footnote{https://www.bbc.com/news/technology-27762088}, where it convinced 33\% of the judges that it was a real human. However, the result remains contentious. More recently, \cite{jones2025people} conducted a study to evaluate the performance of  ELIZA, GPT-3.5, and GPT-4 in a Turing Test setting. Participants engaged in five-minute conversations and were asked to decide whether their conversation partner was human. GPT-4 was mistaken for a human in 54\% of the cases, significantly outperforming ELIZA (22\%) and getting closer to human benchmark (67\%). This deception goes further, as we now see reports that ``GPT-4 told a TaskRabbit worker it was visually impaired to get help solving a CAPTCHA.''\footnote{https://www.businessinsider.com/gpt4-openai-chatgpt-taskrabbit-tricked-solve-captcha-test-2023-3}.

% While the Turing Test seems to be done, Total Turing Test which goes beyond the verbal communication is still not done. One of the aspects of the humans that should be considered is the non-verbal interactions. For the machine to be believable it should have interaction awareness~\cite{bogdanovych2016makes}. Interaction awareness is defined as the ability of an agent that is “to perceive important structural and/or dynamic aspects of an interaction that it observes or that it is itself engaged in”~\cite{dautenhahn2002embodied}. Non-verbal interactions with artificial agents remain challenging to produce convincingly and appropriately~\cite{wang2021examining}.
While the Turing Test may be considered passed, the Total Turing Test~\cite{harnad1991other}, emphasized the need for both linguistic and robotic (nonverbal) capacities.
% which extends beyond verbal communication, remains an open challenge.
Indeed, a key aspect of human behavior that must be addressed is nonverbal interaction. For a machine to appear believable, it must exhibit interaction awareness\cite{bogdanovych2016makes}. Interaction awareness is defined as an agent's ability ``to perceive important structural and/or dynamic aspects of an interaction that it observes or that it is itself engaged in''\cite{dautenhahn2002embodied}. Generating convincing and context-appropriate nonverbal interactions in artificial agents continues to be a significant challenge~\cite{wang2021examining}.

In this paper, we propose a new question: ``Can machines react?'' Building on Turing’s original framework, we introduce a test in which a human judge interacts with an embodied agent for 1-minute, and decides whether it was teleoperated or autonomous. To pass the test, the autonomous agent must convincingly mimic the behavior of a tele-operated agent, leading the judge to misidentify it as a human-controlled one.
This research builds on prior work examining awareness, nonverbal communication, and affective behaviors in interactions with artificial agents. Our focus is on a subset of socially interactive agents~\cite{10.1145/3477322}, including social robots and interactive virtual agents (IVAs) that utilize visual sensing technologies such as cameras. Throughout this paper, we refer to these systems as interactive characters. The main contributions of this paper are:
\begin{enumerate}
\item We propose a task to answer ``Can machines react?'', the React to This (RTT) test designed specifically for nonverbal behaviors
\item We present results from an initial experiment designed to assess how autonomous interactive characters can convince humans that they are tele-operated in real-time social interactions.
\end{enumerate}

\begin{figure}[t]
    % \centering
    \includegraphics[width = 0.49 \textwidth]{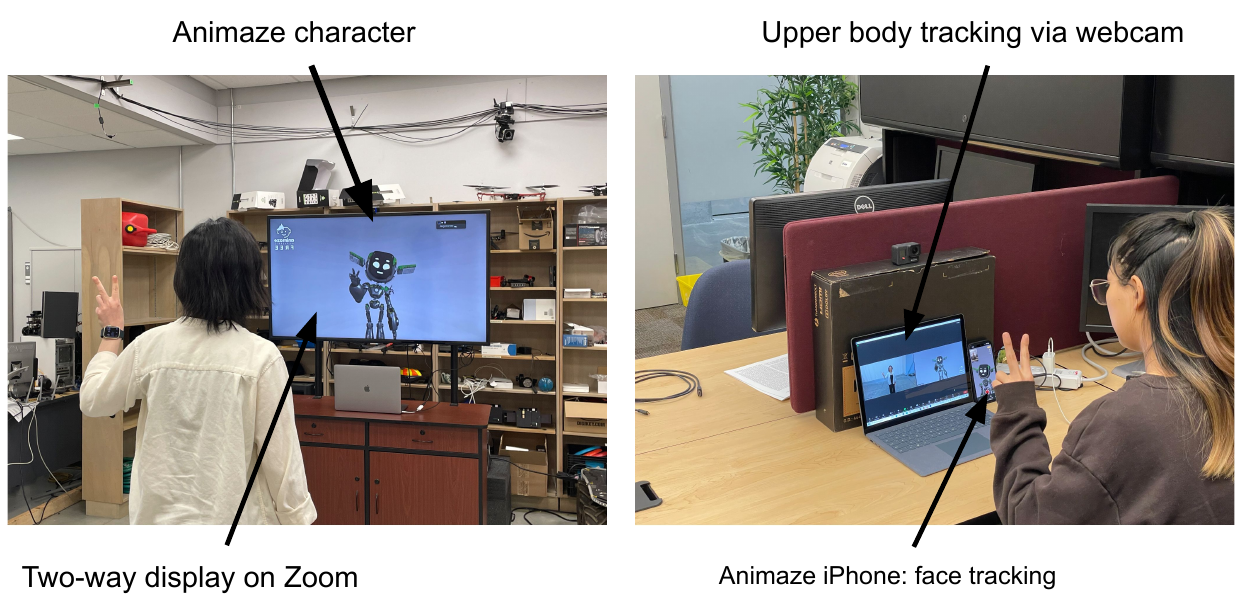}
    \vspace{-5mm}
    \caption{Physical setup of our study: participant interacting with virtual character (left), and teleoperator using face and upper body tracking to control the virtual character (right).}
    \vspace{-8mm}    
    \label{fig:setup}    
    % \Description{Two photos capture the physical setup of the study. The photo on the left shows a participant interacting with a virtual character displayed on a smart screen. The photo on the right shows the teleoperator remotely controlling the virtual character from another room with an iPhone face tracker, and the webcam on a laptop.}
\end{figure}

\section{Related Work}
%Short paragraph 1: A sentence about related areas

\emph{Turing Test.} The original Turing Test~\cite{turing2009computing} was introduced to explore the question: "Can machines think?" Since the answer depends heavily on how the terms 'machine' and 'think' are defined, Turing reframed the question using a game called the Imitation Game, involving a human, a machine, and a human interrogator. In this game, the interrogator communicates with the other two through a text-based interface and the goal is to identify which of the other two participants is the human and which is the machine. For the machine to succeed, it must convince the interrogator that it is the human. Subsequently, other tests of intelligence have been proposed, such as the Lovelace~\cite{bringsjord2003creativity} and its successor, Lovelace 2.0~\cite{riedl2014lovelace} tests, which emphasize creativity as a core component of intelligence.

\emph{Total Turing Test.} While the original test only assessed linguistic ability, Searle~\cite{searle1980minds} argued that even a mindless symbol manipulator could pass it. Therefore, to truly evaluate whether a system might have a mind, it must go beyond verbal communication and exhibit full human-like interaction with the physical world, implying the candidate should be a robot, not just a text-based system. The Total Turing Test (TTT)~\cite{harnad1991other} extends Turing’s original language-only test by requiring a candidate to replicate everything a real person can do in the physical world, encompassing both linguistic and robotic abilities. Unlike Turing’s version, the TTT emphasizes that our bodily interactions with the world through a wide range of nonverbal behaviors are just as essential as our linguistic capacities. 

\emph{Nonverbal Turing Test.} Ciardo et al. \cite{ciardo2022human} demonstrated that the standard deviation and distribution shape of reaction times influence humans' ability to distinguish robots from humans. Similarly, \cite{pfeiffer2011non} conducted a nonverbal Turing test using an eye-tracking paradigm to investigate how gaze behavior affects the perception of humanness in virtual agents. \cite{ventrella2010gestural} proposed nonverbal version of the Turing Test based on gesture. They developed two gestural AI systems that mimicked human movement through head and hand positions and evaluated their believability by asking participants to judge whether the gestures were human or machine-generated. Additional studies \cite{gurion2018real, hingston2009turing} have also explored nonverbal Turing Tests in gaming environments.

\emph{Consciousness and Interaction awareness}. Based on the movie Ex Machina, Shanahan proposed the Garland Test~\cite{shanahan2024simulacra}, where a human ``ascribes consciousness to an AI system even though they know that it is an artefact". In humans, consciousness is tested using, for instance, the Glasgow Coma Scale~\cite{STERNBACH200067}, which tests eye, basic verbal and bodily motor responses to sound, painful stimuli and movement commands. Relatedly, interaction awareness involves the ability to perceive dynamic aspects of an interaction~\cite{dautenhahn2002embodied}. Specifically, Dautenhahn et al. suggest that an ``important ability of an interaction-aware agent is to \emph{track, identify, and interpret visual interactive behavior}"~\cite{dautenhahn2002embodied}, along the continuum of interaction formality~\cite{hutchby2008conversation}. This includes informal interactions such as play, semi-formal interactions such as greetings, and very formal interactions such as scripted law proceedings. As an example, Aldebaran Robotics\footnote{https://www.aldebaran.com/} proposed the Basic Awareness module on their NAO and Pepper robots, which includes tracking detected humans and looking in the direction of stimuli such as movement or sound, towards the illusion of life.

\begin{figure}[t]
    % \centering
    \includegraphics[width = 0.5 \textwidth]{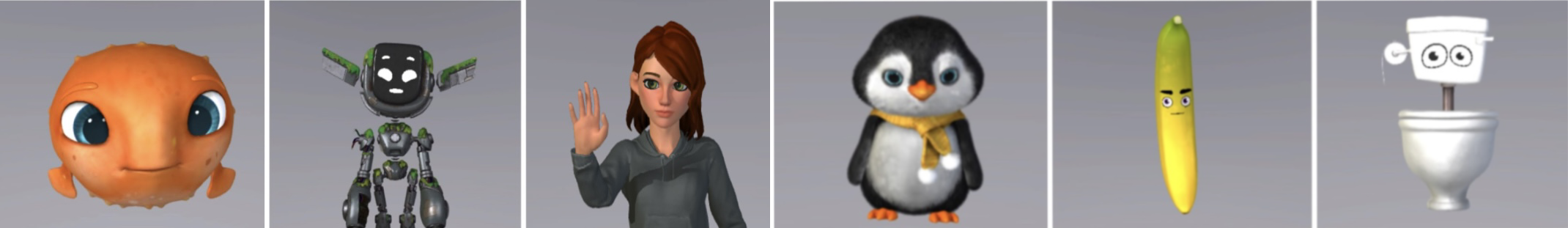}
    \vspace{-5mm}    
    \caption{Participants were asked to test 6 interactive virtual characters physically, emotionally, and socially.}
    \vspace{-7mm}    
    \label{fig:character}    
    % \Description{Two photos capture the physical setup of the study. The photo on the left shows a participant interacting with a virtual character displayed on a smart screen. The photo on the right shows the teleoperator remotely controlling the virtual character from another room with an iPhone face tracker, and the webcam on a laptop.}
\end{figure}

\section{React to This (RTT) Test}
We propose the React to This (RTT) test, a nonverbal Turing test for embodied AI agents. In this test, humans are given 1 minute to interact nonverbally with an embodied, anthropomorphic agent. At the end, the human should state whether they believe the agent was teleoperated or autonomous. The AI agent passes the test if the majority of humans believe it was teleoperated.

\section{Experiments}
We conducted an initial trial of the React to This (RTT) test to explore the interactive behaviors people performed to test the awareness and reactiveness of the intelligent agents given 1 minute of interaction time (Fig.~\ref{fig:setup}, left). In \cite{zhang2024react}, we reported what people do to test a character physically,  emotionally, and socially. In this paper, we aim to provide insights into the kinds of nonverbal abilities that were tested, as well as results and caveats of this initial RTT test.

\subsection{Setup and Materials}
We prepared a Wizard-of-Oz (WoZ) experimental setup (Fig.~\ref{fig:setup}) using three off-the-shelf software packages to create interactions between human participants and virtual characters. 1. \emph{Animaze}\footnote{https://www.animaze.us/} provided ready-to-use animated characters and built-in head pose and facial expression tracking. 2. \emph{Webcam Motion Capture}\footnote{https://webcammotioncapture.info/}  tracked and mapped upper body movements to the characters. 3. \emph{Zoom}\footnote{https://zoom.us/} enabled real-time, bi-directional interaction. On the participant side, the virtual character was displayed on a large TV screen (55 inches) oriented horizontally. Participants were asked to stand approximately 1.5m from the display.

\subsection{Study Design and Task}
\label{subsection: study design}
We conducted a WoZ study with 20 adults (gender: 9/10 women/men, 1 prefer not to disclose; age: $28.5\pm12.98$) participants interacting with 6 different virtual characters (Fig. \ref{fig:character}) for 1 minute each. The study was approved by the university ethics board. To minimize uncertainty regarding participants’ assumptions about character agency, we informed them that all characters were autonomous. We also told them that the characters could not process audio input and did not understand human language. As a result, the study was entirely nonverbal. Participants were given one minute, and were told ``Your goal is to test what the character can and cannot do physically, emotionally, and socially". After the interaction, participants were interviewed about their perception of the characters' autonomy (e.g., whether they believed the characters were truly autonomous, and why). They were then debriefed and informed that the characters had been teleoperated. Participants could then revise their consent for data use, if desired.

\subsection{Data Collection and Analysis}

We used ELAN \cite{ELAN} to annotate video recordings of the sessions. Each video was independently reviewed by two annotators. The primary annotator segmented and labeled the footage, while the secondary annotator reviewed the annotations and noted any potential discrepancies. We used a discuss-until-consensus approach common in qualitative research \cite{harding2013analysing}. In cases of persistent disagreement, a third annotator was consulted. We then conducted behavior classification using thematic analysis. The human interactive behavior codebook and more details can be found in \cite{zhang2024react}. The full set of annotations and anonymized videos can be downloaded from \url{https://rosielab.github.io/react-to-this/}.

\section{Results}

\begin{figure}
    % \centering
    \includegraphics[width = 0.49\textwidth]{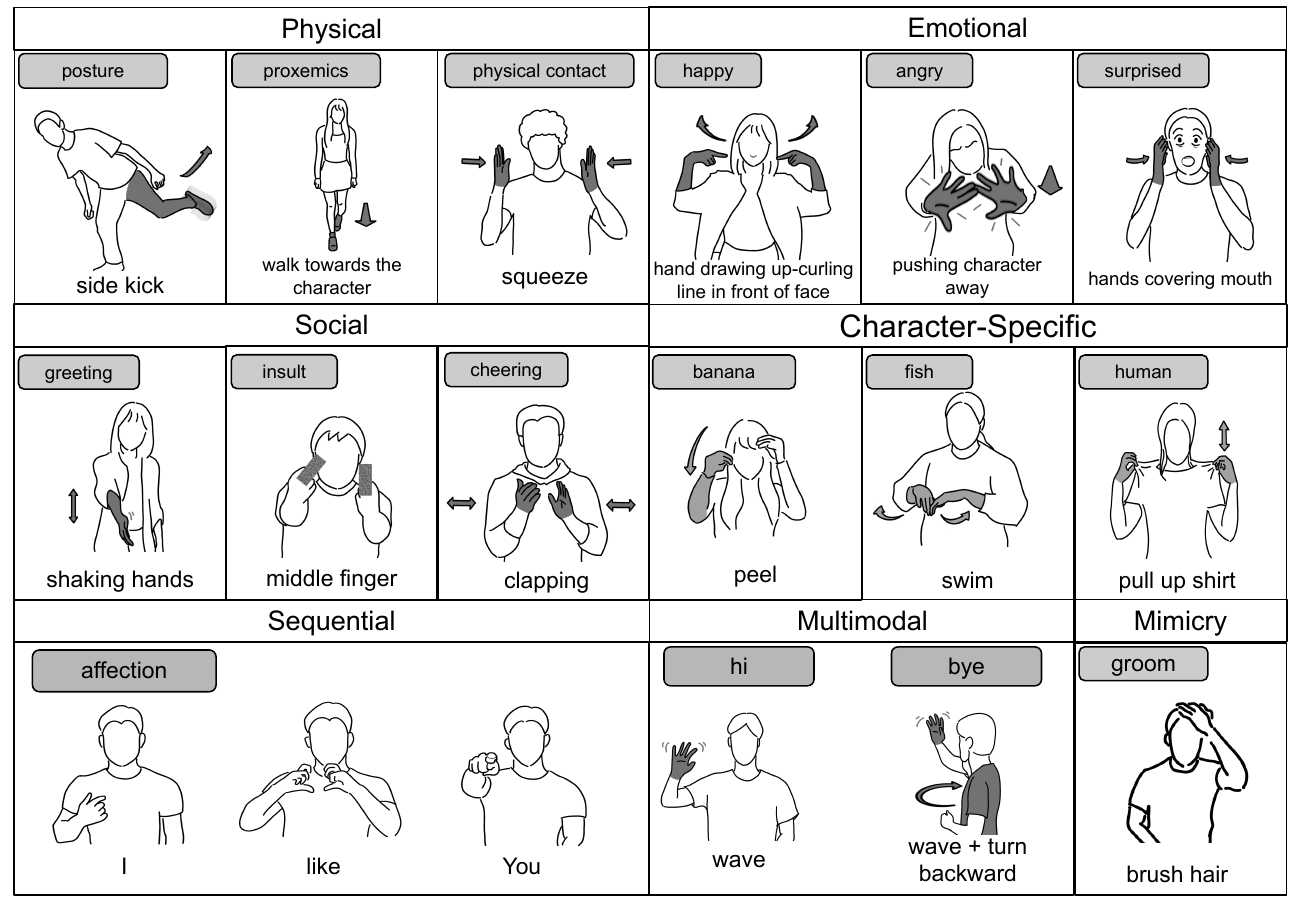}
    \vspace{-5mm}
    \caption{Example nonverbal behaviors used to test agents}
    \vspace{-7mm}
    
    \label{fig: examples}

\end{figure}

A total of 1,169 interactions were recorded during the study, including 1,111 participant-initiated interactions and 58 agent-initiated interactions. Of the 1,111 participant-initiated interactions, 999 were character-agnostic. In this section, we highlight the major and most interesting findings from our observations (see Fig. \ref{fig: examples}).

\emph{Physical testing} refers to participants' propensity to examine the character’s motor and sensing capabilities. For motor testing, participants changed their posture (e.g., bending, raising their arm or leg), presumably to see if the agent could copy them. In terms of sensing, they performed proxemic changes (e.g., moving forward and side to side), appearing to see if the agent could visually track their motion. They also tried to poke or squeeze the agents using pretend physical contact, again seeming to seek a reaction. These tests are reminiscent of some aspects of the Glasgow Coma Scale (e.g. asking the patient to copy their motor movement and reactions to poking, and seeking to assess their eyes.)

\emph{Emotional testing} was performed by exhibiting various emotions with facial expressions and gestures (Fig. \ref{fig: examples}, top right). Notably, participants often used pantomime actions to exaggerate emotional expressions, especially for characters such as the penguin, suggesting low expectations of the character’s emotional intelligence. A wide diversity of emotional behaviors was observed, which poses a challenge for emotion expression recognition systems built into interactive agents.

\emph{Social testing} was observed. Notably, culture-specific behaviors, such as namaste, were also performed by the participants during the study. This suggests the need for intelligent agents to interpret and respond to socially and culturally grounded nonverbal behaviors.  We also noticed aggressive behaviors, such as punching or firing handguns. This is reminiscent of HRI studies such as Kanda et al. reporting children's aggressive behaviors toward robots in shopping malls~\cite{nomura2015children}.

\emph{Identity testing and context-awareness} were assessed by character-specific behaviors (Fig. \ref{fig: examples}, middle right). For instance, participants tested characters’ physical affordances (e.g., trying to peel a banana), engaged in role-based interactions (e.g., dancing with a robot), or referred to visual accessories (e.g., complimenting the penguin’s scarf).

\emph{Mimicry}\cite{chartrand1999chameleon} was observed in both conscious and unconscious forms during the study. Inspired by the characters' reactive behaviors, participants sometimes intentionally repeated the characters' behavior to interact playfully. Participants also unconsciously mimicked the characters' behaviors, such as grooming (Fig. \ref{fig: examples}, bottom right).

\emph{Comprehension of sequential and multimodal behaviors} was tested when a single action was not enough for the participants to express themselves. Sequential behaviors, such as ``I like you" were composed of 3 actions: 2 pointing/diectic and 1 heart shape hand gestures. The same gesture could also convey different meanings; we differentiated ``I don't know" and ``tada" shrugs by examining the facial expressions from the participants  (Fig. \ref{fig: examples}, bottom).

\emph{Teleoperated or Not?}
After the study, we interviewed the participants to see if they thought the characters were autonomous as told beforehand, or if they suspected that they were teleoperated. 8/20 participants suspected that at least one character was teleoperated. 9/20 thought the characters were fully autonomous. One participant was reluctant to give a clear answer to this question, and one answer was omitted due to video corruption. Some explanations were provided.

\emph{Reasoning about Autonomy}. Participant 114 (P144) reported themself to be a member of the otaku community with extensive experience with virtual YouTubers. They suspected that guessing the control of the character was a hidden task. The human character was voted for being teleoperated by 4 participants, for being the most reactive, and having a short response time. The penguin and the fish were considered to be teleoperated for the same reasons by 2 participants. Interestingly, P114 reported that the robot was the least likely to be teleoperated for its limited reactions and overall poor performance during the interaction. In addition, the two inanimate object characters (banana and toilet) were not recognized as teleoperated by anyone. It is possible that the low expectations for these characters allowed them to be considered autonomous more easily.

 Among the participants who did not think any of the characters were teleoperated, 2 of them thought that any interaction between a human and a computer had to be pre-programmed and not involve a human's input. 3/19 participants stated that they had higher expectations of teleoperated characters, and ours did not reach that expectation. One participant reported perceiving some of the characters' behavior as unnatural and hence not controlled by humans. One participant claimed that all characters seemed to have uniform reaction times, which appeared preprogrammed. One participant believed that the short reaction time made the characters more likely to be autonomous.

\section{Takeaways for Test Design}
Based on these results, we summarize design suggestions to effectively implement the RTT nonverbal Turing test, and suggestions for agents that aim to pass the test.

\textbf{Latency} can significantly affect how users perceive the agent’s intelligence mentioned by 4 participants. In our study, latency was periodically caused by network issues. In these cases, the delay was too long to be attributed to the agent’s inner processing of the stimuli or the action being performed. Two participants in our study reported that shorter reaction times made the characters appear more intelligent and reactive. Therefore, an implementation of the test should aim to eliminate network delays, and agents should manage timing effectively (neither too long nor too short reaction times, and not too uniform).

\textbf{Morphology} influences how humans evaluate their intelligence by altering their expectations. Comparisons should ideally be drawn between agents of the same visual appearance, or at least of the same type (e.g., human or robot-like, zoomorphic, inanimate objects). Our study underscored that character-specific behaviors were shaped by identity and morphology: participants tended to perform more cognitively challenging tasks with the human character, expecting higher intelligence. The ability to interact with accessories also contributed to perceived awareness. For example, participants attempted to interact with the scarf on the penguin and the hoodie on the human character. Therefore, object affordances embedded in the agent's physical design, clothing, or environmental surroundings must also be considered, as they influence user expectations and behavior.

\textbf{Sequential and multimodal behaviors} are useful for evaluating the agent’s understanding of temporal structure and its memory capacity. However, increasing behavioral complexity also extends the duration and difficulty of the test. Believability breaks down quickly if the agent cannot preserve relevant interaction history or correctly segment and respond to complex behavior sequences. 
    
\textbf{Interface} plays an important role in the test. In our setup, participants interacted with virtual characters via Zoom, with the characters displayed on a screen. This interface constrained the types of tests participants could perform, particularly by preventing exploration of physical contact. With physically embodied agents, we would expect more complex and challenging behaviors, especially involving haptics. Thus, the modality of the interface plays a crucial role in shaping the test content and participants' strategies.

Designing a non-verbal Turing test requires careful consideration of factors that may unintentionally influence participant perception. Uncontrolled variables, such as latency, morphology, and interface limitations should be minimized or systematically accounted for in the test design.

% \section{Discussion}
% We considered other options such as providing 8 different agents, and asking particants to test them nonverbally to find out which one was teleoperated. We also believe that if agents can pass the 1-minute version of the test, that longer versions could also be trialed. Indeed, pet robots are meant to be believable companions that. In addition, we can propose varied morphologies of the test. It appeared that the human 

% \input{Sections/discussion}

\section{Discussion and Future Work}
While our proposed test provides a new lens through which to evaluate machine believability via nonverbal interaction, it is only a first step toward understanding the full spectrum of social responsiveness in artificial agents. 
It is worth noting that we used virtual characters rather than physical robots. As a result, the test does not fully assess physical robotic capabilities or object interactions as in the Total Turing Test (e.g. pouring water) and may elicit different responses compared to interactions with physically embodied agents. Three-dimensional interaction is also lacking, e.g. moving around the agent. Future work can explore using actual robots instead of a screen, to better evaluate how physical interaction affects believability.
Furthermore, in our experiment, we only used human-controlled characters but informed participants that the characters were autonomous. This may have influenced their judgments. Future experiments should include both teleoperated and autonomous characters, with participants asked to distinguish between them. This test is also limited in that teleoperators typically do not have connections to physical interaction, and therefore would not experience real reactions such as pain in response to aggression from the agent.
Our study was conducted at a North American post-secondary institution, which may limit the generalizability of our findings across broader populations. Cultural background, age, and neurodiversity \cite{ZhangChuxuan2024Rtt} can all influence how people perceive and interact with artificial agents. Future work should aim to include more diverse participant samples to better capture this variability. Additionally, our study excluded other nonverbal modalities such as auditory cues (e.g., clapping), which future research could incorporate to provide a more comprehensive assessment. Finally, the RTT test mainly focuses on visual and behavioral reactions, isolated from complex language. Future work can combine this nonverbal test with speech, even basic language similar to the Glasgow Coma Scale, to move autonomy toward grander goals such as the Total Turing Test.

\section{Conclusion}
In this paper, we propose the React to This (RTT) test, which asks the question “Can machines react?” as a nonverbal Turing Test. To explore this idea, we conducted an experiment with six different virtual agents, asking participants to interact with each for one minute. We asked participants whether they believed each character was pre-programmed or controlled by a human, and recorded their responses and reasoning. We also provide design takeaways for implementations of the RTT test and describe non-verbal behaviors necessary for a character to be perceived as believable.

% In this study, we discovered 188 unique actions and 51 socio-emotional behavior categories among 1169 non-verbal interactions between participants and 6 virtual characters, contributing to the list of target classes for visual gesture recognition algorithms. With a bottom-up analysis method, we created a rich and diverse behavior codebook to guide designers and programmers of interactive agents/robots. The 188 actions and corresponding meanings also could help provide a list of classes for machine learning gesture recognition algorithms to target. The set of abundant interactive behaviors in our codebook can be applied to interactive agents deployed in various settings (e.g. video games, theme park, education) in the future.

\bibliographystyle{IEEEtran}
\bibliography{custom.bib}

\end{document}